\documentclass[letterpaper, 10pt, oneside, journal, final, twocolumn]{IEEEtran}
\usepackage[latin1]{inputenc}
\usepackage[pdftex]{graphicx}
\usepackage{array, multirow}
\usepackage{color}
\usepackage{booktabs}
\usepackage{tikz, pgfplots}
\usetikzlibrary{arrows.meta}
\usetikzlibrary{spy}
\usepackage{amssymb}
\usepackage{amsmath}
\usepackage{tabularx}
\usepackage{makecell}
\usepackage{caption}
\usepackage{subcaption}
\graphicspath{{figures/}}
\pgfplotsset{compat=1.16}
\def\BibTeX{{\rm B\kern-.05em{\sc i\kern-.025em b}\kern-.08em
		T\kern-.1667em\lower.7ex\hbox{E}\kern-.125emX}}

\usepackage[nolist]{acronym}
\begin{acronym}   
	\acro{CSS}{chirp spread spectrum}
	\acro{SER}{symbol error ratio}
	\acro{SNR}{signal-to-noise ratio}
	\acro{LoRa}{Long-Range}
	\acro{IoT}{Internet of things}
	\acro{FSK}{frequency shift keying}
	\acro{AWGN}{additive white Gaussian noise}
	\acro{LPWAN}{low power wide area network}
	\acro{PHY}{physical layer}
	\acro{SE}{spectral efficiency}
	\acro{EE}{energy efficiency}
	\acro{MLD}{maximum likelihood detection}
	\acro{DSSS}{direct sequence}
	\acro{FHSS}{frequency hoping}
	\acro{FFT}{fast Fourier transform}
	\acro{SF}{spreading factor}
	\acro{IQCSS}{in-phase and quadrature chirp spread spectrum}
	\acro{MLE}{maximum likelihood estimator}
	\acro{DFT}{discrete Fourier transform}
	\acro{SFD}{start frame delimeter}
	\acro{GSM}{Global System for Mobile Communications}
	\acro{DCRK-CSS}{discrete chirp rate keying CSS}
	\acro{CP}{cyclic prefix}
	\acro{LS}{least squares}
\end{acronym}

\begin{document}
\IEEEoverridecommandlockouts
\title{Alternative Chirp Spread Spectrum Techniques for LPWANs}

\author{Ivo Bizon Franco de Almeida\IEEEauthorrefmark{1}, Marwa Chafii\IEEEauthorrefmark{2}, Ahmad Nimr\IEEEauthorrefmark{1} and Gerhard Fettweis\IEEEauthorrefmark{1}
	
\thanks{\IEEEauthorrefmark{1}Vodafone Chair Mobile Communications Systems, Technische Universit{\"a}t Dresden (TUD), Germany. \IEEEauthorblockA{\IEEEauthorrefmark{2}ETIS, UMR8051, CY Cergy Paris Universit{\'e}, ENSEA, CNRS, France}. Emails: \{ivo.bizon, ahmad.nimr, gerhard.fettweis\}@ifn.et.tu-dresden.de, marwa.chafii@ensea.fr.}}

\maketitle
\begin{abstract}
	~ Chirp spread spectrum (CSS) is the modulation technique currently employed by Long-Range (LoRa), which is one of the most prominent Internet of things wireless communications standards. The LoRa physical layer (PHY) employs CSS on top of a variant of frequency shift keying, and non-coherent detection is employed at the receiver. While it offers a good trade-off among coverage, data rate and device simplicity, its maximum achievable data rate is still a limiting factor for some applications. Moreover, the current LoRa standard does not fully exploit the CSS generic case, i.e., when data to be transmitted is encoded in different waveform parameters. Therefore, the goal of this paper is to investigate the performance of CSS while exploring different parameter settings aiming to increase the maximum achievable throughput, and hence increase spectral efficiency. Moreover, coherent and non-coherent reception algorithm design is presented under the framework of maximum likelihood estimation. For the practical receiver design, the formulation of a channel estimation technique is also presented. The performance evaluation of the different variants of CSS  is carried out by inspection of the symbol error ratio as a function of the signal-to-noise ratio together with the maximum achievable throughput each scheme can achieve.
\end{abstract}

\acresetall

\section{Introduction} \label{introduction}

\IEEEPARstart{R}{ecently}, a lot of attention has been drawn towards long-range and low-power consuming wireless communication schemes \cite{lpwans}. 
\ac{LoRa} is a wireless communication protocol that has got preference among the schemes considered primary for \ac{IoT} applications \cite{survey1}. 
The main application of \ac{LoRa}, and \acp{LPWAN} in general, is to provide connectivity for mobile and stationary wireless end-devices that require data rates in the order of tens of kbps up to a few Mbps within a coverage area up to tens of kilometers. 
For maximization battery life and minimizing end-device costs, low energy consumption and simple transceiver design are also desirable characteristics for \acp{LPWAN} devices \cite{iot_citation}.

The \ac{PHY} of \ac{LoRa} has gained considerable attention within the academic community, and several papers have been published with investigations on the characteristics of \ac{LoRa} \ac{PHY} and MAC schemes \cite{lora_limits,lora_capacity}.
Its modulation scheme is based on \ac{CSS} in conjunction with \ac{FSK}.
Some authors have proposed enhancements to the \ac{LoRa} \ac{PHY} framework.
For instance, encoding extra information bits on the phase of the spreading chirp waveform has been proposed in \cite{psk_lora}, and similar work employs pulse shaping on top of the chirp waveform to reduce the guard-band, and thus increases the number of channels within the available frequency band \cite{efficient_css}.
Most recently, an approach that uses the chirp rate to carry one extra information bit  has been proposed in \cite{ssk}.
A combination of pulse position modulation with \ac{CSS} has been proposed in \cite{suggestion_1}, where the orthogonality among time shifted versions of the chirp signal is exploited to allow multiple access in the downlink.
Moreover, the analysis of inter-spreading factor in \ac{LoRa} systems has been investigated in \cite{suggestion_2} and \cite{lora_sf_orthogonality}, showing that the symbol error rate increases significantly in high signal-to-interference ratios.
The robustness against interference of the schemes presented here can be investigated in the future.
While in \cite{suggestion_3}, the performance of the \ac{LoRa} \ac{PHY} is compared against a frequency hopping technique, and results present that \ac{LoRa} has a higher transmission success probability for small data packets.
Therefore, inspired by the aforementioned works, and departing from the \ac{LoRa} \ac{PHY} framework, we propose a more broad scheme that encodes multiple information bits on the discrete chirp rate.
The scheme proposed in \cite{iqcss} is revisited here, and while it employs coherent detection, the new scheme that exploits the chirp rate as information carrier is able to benefit from non-coherent detection when operating over flat-fading channels.
The main goal of this paper is to present two innovative modulation designs that improve the current \ac{LoRa} \ac{PHY}, while one of them is build upon a coherent receiver, the other is build based on a non-coherent receiver.
Furthermore, maximum likelihood estimation for the transmitted data symbols using the classical \ac{CSS} transceiver is presented.

The main contributions contained in this paper are:
\begin{itemize}
	\item A detailed transceiver description of two proposed modulation techniques based on \ac{CSS}, namely \ac{IQCSS} and \ac{DCRK-CSS}, which are able to increase \ac{SE} and \ac{EE} when compared to the conventional \ac{LoRa} \ac{PHY} scheme.
	
	\item A mathematical derivation of the optimum receiver for \ac{CSS} under the framework of maximum likelihood estimation.
	
	\item A low complexity channel estimation technique that can be used with the current \ac{LoRa} \ac{PHY} data packet structure, thus allowing coherent reception.
\end{itemize}

%

The remainder of the paper is organized as follows: 
Section \ref{subsec:chirp_signal} presents the mathematical foundations of discrete-time chirp signals from its continuous-time definition, while Sections \ref{subsec:fskcss} and \ref{subsec:mld} present the transmission and reception of \ac{CSS} modulation under the framework of maximum likelihood estimation. 
Section \ref{sec:alternative} contains a detailed description of the transceivers which employ variants of \ac{CSS}.
Firstly in \ref{subsec:lora}, the \ac{LoRa} \ac{PHY} standard \cite{loradatasheet}, subsequently in \ref{subsec:iqcss} the recently proposed \ac{IQCSS} \cite{iqcss}, and lastly in \ref{subsec:dcrk} the proposed \ac{DCRK-CSS} are presented.
Section \ref{sec:performance} investigates the performance of the proposed schemes under different wireless channel models.
Finally, the paper is concluded in Section \ref{sec:conclusions} with future insights, and directions on the topic of \acp{LPWAN}.
\section{Chirp Spread Spectrum Maximum Likelihood Estimation} \label{sec:mle_css}

\subsection{The Chirp Signal} \label{subsec:chirp_signal}
This subsection aims to give an analytical description of the chirp signal.
In \ac{CSS}, as well as in other spread spectrum techniques, such as \ac{DSSS} and \ac{FHSS}, the information is transmitted  using a bandwidth much larger than required for a given data rate, i.e., the information signal is spread over the bandwidth to benefit from spreading gain. 
Particularly in \ac{CSS}, multiplication by a chirp signal is responsible for the energy spreading in frequency \cite{css1, theory_chirp}. 
The linear-chirp refers to the frequency variation of the signal, which increases linearly with time. 

The chirp waveform can be described by
\begin{equation} \label{chirp_continouos_time}
	c(t) = \begin{cases}
		\exp\left( j\varphi(t) \right) & \text{for } -T/2 \leq t \leq T/2 \\
		0 & \text{otherwise,}
	\end{cases}
\end{equation}
where $\varphi(t) = \pi (at^2 + 2bt)$, i.e., a quadratic function of time. 
The signal instantaneous frequency is defined as 
\begin{equation}
	v(t) = \frac{1}{2\pi} \frac{d \varphi(t)}{dt} = at + b,
\end{equation}
which shows that the frequency varies linearly with time. 
Moreover, the chirp rate is defined as the second derivative of $\varphi(t)$ w.r.t. $t$ as
\begin{equation}
	u(t) = \frac{1}{2\pi} \frac{d^2 \varphi(t)}{dt^2} = a.
\end{equation}
The up-chirp corresponds to the case where $u(t)>0$, and the down-chirp when $u(t)<0$.

Let $B$ (Hz) represent the bandwidth occupied by the chirp signal.
The signal frequency varies linearly between $-B/2 + b$ and $B/2 + b$ within the time duration $T$ (s).
If the term $b=0$ and $a \neq 0$, the resulting waveform is the \textit{raw} chirp with starting frequency $-B/2$ and end frequency $B/2$. 
Conversely, if $a=0$, the complex exponential is obtained.
Sampling \eqref{chirp_continouos_time} at the Nyquist rate, i.e., $T_s = 1/B$, where $T_s$ (s) is sampling time interval, the baseband discrete-time chirp signal is given by
\begin{equation}
	c(nT_s) = \begin{cases}
		\exp\left( j\varphi(nT_s) \right) & \text{for } n = 0, \ldots, \, N-1 \\
		0 & \text{otherwise,}
	\end{cases}
\end{equation}
where $N = T/T_s$ is the total number of samples within $T$ seconds. Setting $b=0$, and $a = MB/T$, the discrete-time raw up-chirp becomes
\begin{equation}
	c[n] = \exp\left( j\pi M n^2/N \right),
\end{equation}
where $M$ represents the discrete chirp rate.

\subsection{\ac{CSS} Modulation} \label{subsec:fskcss}

At the transmitter side, the discrete-time chirp signal $c[n]$ is used for spreading the information signal within the bandwidth $B$ via multiplication.
The transmit signal can be described as 
\begin{equation} \label{tx_signal_LoRa}
	x_{k}[n] = \sqrt{\frac{E_s}{N}} \exp \left( j\frac{2\pi }{N}kn \right) c[n],
\end{equation}
where the complex exponential term has its frequency depending upon the data symbol $k$, which is obtained from a bit-word $\mathbf{b} \in \left\lbrace 0,\,1 \right\rbrace^{\mathrm{SF}}$ as
\begin{equation}
	k = \sum_{i=0}^{\mathrm{SF}-1} 2^i \left[\mathbf{b}\right]_i,
\end{equation}
where the \ac{SF} represents the amount of bits in each bit-word.
Note that each waveform has $N = 2^{\mathrm{SF}}$ samples for having distinguishable waveforms. 
The data symbols are integer values from the set $ \mathbb{K} = \left\lbrace 0, \ldots, \, 2^{\mathrm{SF}} - 1\right\rbrace $, which contains $N$ elements. 
$E_s$ represents the signal energy. 

To explore alternative ways of generating the transmit signal, let us analyze the period of the raw up-chirp, and for the sake of brevity let us assume $M=1$.
Then, 
\begin{equation}
	\exp\left( j\pi \frac{\left(n + N\right)^2}{N}\right) = \exp\left(j\pi n^2/N\right) \exp\left( j2\pi n\right) \exp\left( j\pi N\right),
\end{equation}
where the terms $\exp\left( j2\pi n\right)$ and $\exp\left( j\pi N\right)$ equal the unity for all $n \in \mathbb{K}$ and $N \in \left\lbrace 2^{6},\,\ldots,\, 2^{12} \right\rbrace$.
Therefore, the raw up-chirp has period $N$, and the \ac{LoRa} transmit signal can be readily obtained via a discrete circular time shift operation.
Hence, (\ref{tx_signal_LoRa}) can also be written as
\begin{equation}
	x_{k}[n] = \sqrt{\frac{E_s}{N}} c[n+k]_N,
\end{equation}
where $\left[\cdot\right]_N$ represents the circular shift.
This equation presents simplified generation complexity when compared to (\ref{tx_signal_LoRa}), since it is necessary to just store one vector corresponding to the raw up-chirp and read it circularly.

\subsection{Maximum Likelihood Estimation of \ac{CSS}} \label{subsec:mld}

The discrete-time complex-valued received signal after synchronization can be described as
\begin{equation} \label{rx_signal_LoRa}
	y[n] = f_h\left(x_k[n]\right) + w[n],
\end{equation}
where $f_h\left(\cdot\right)$ represents the equivalent channel function, and $w[n]$ is assumed to be \ac{AWGN} with zero mean and $\sigma^2_w$ variance.

Assuming knowledge of the equivalent channel function, the \ac{MLE} for the transmitted data symbol $k$ is given as
\begin{equation} \label{mle}
	\hat{k} = \arg\min_{k \, \in \, \mathbb{K}} \sum_{n=0}^{N-1} \big| y[n] - f_h\left(x_k[n]\right) \big| ^{2}.
\end{equation} 

\subsubsection{AWGN Channel}

Under pure \ac{AWGN} channel the equivalent function is the identity, i.e., $f_h\left(x_k[n]\right) = x_k[n]$, and
\begin{equation} 
	\big|y[n] - x_k[n]\big|^{2} = \big|y[n]\big|^2 + \big|x_k[n]\big|^{2} - 2\, \Re\left\lbrace x_k^*[n]y[n] \right\rbrace,
\end{equation}
where $(\cdot)^*$ and $\Re\left\lbrace\cdot\right\rbrace$ denote the conjugate and real part extraction operators, respectively, and note that $\big|x_k[n]\big|^{2} = E_s/N \; \forall \; n \, \in \, \mathbb{K}$.
Consequently, (\ref{mle}) can be modified to 
\begin{equation}
\begin{split} \label{mle2}
	\hat{k} & = \arg\max_{k \, \in \, \mathbb{K}} \, \sum_{n=0}^{N-1} \Re\big\lbrace x_k^*[n]y[n] \big\rbrace \\
            & = \arg\max_{k \, \in \, \mathbb{K}} \, \sum_{n=0}^{N-1} \Re\left\lbrace \sqrt{\frac{E_s}{N}} \exp \left( -j\frac{2\pi }{N}kn \right) c^*[n]y[n] \right\rbrace \\
            & = \arg\max_{k \, \in \, \mathbb{K}} \, \Re\left\lbrace \sum_{n=0}^{N-1} \exp \left( -j\frac{2\pi }{N}kn \right) c^*[n]y[n] \right\rbrace.
\end{split}
\end{equation}
By inspecting the maximization problem in (\ref{mle2}), one can see that the demodulation process can be efficiently executed by firstly multiplying the received signal with the conjugated raw up-chirp, hereafter named down-chirp, secondly computing the \ac{DFT} of this result, and thirdly selecting the frequency bin that maximizes the real part of the compound result. 
Luckily, this operation can be efficiently carried via the \ac{FFT} algorithm.
Therefore, the \ac{MLE} for the transmitted data symbol is given by
\begin{equation} \label{mle_awgn}
	\hat{k} = \arg\max_{k \, \in \, \mathbb{K}} \, \Re\left\lbrace R(k) \right\rbrace,
\end{equation}
where $R(k) = \mathcal{F}\left\lbrace r[n]\right\rbrace$, $r[n] = c^*[n]y[n]$, and $\mathcal{F}\left\lbrace \cdot \right\rbrace$ the \ac{DFT} operator. 

\subsubsection{Flat-fading Channel}

Assuming a flat fading wireless channel, the equivalent channel function becomes $f_h\left(x_k[n]\right) = h x_k[n]$, where $h$ denotes a complex-valued gain.
If a channel estimation mechanism is available at the receiver side, coherent detection can be employed, and by following the same path that led to (\ref{mle_awgn}), the estimated data symbol can be obtained as
\begin{equation} \label{mle_flat}
	\hat{k} = \arg\max_{k \, \in \, \mathbb{K}} \, \Re\left\lbrace \tilde{R}(k) \right\rbrace,
\end{equation}
where $\tilde{R}(k) = \mathcal{F}\left\lbrace \tilde{r}[n]\right\rbrace$, $\tilde{r}[n] = h^*c^*[n]y[n]$.
However, if non-coherent detection is implemented at the receiver, the random phase rotation induced by the complex channel gain cannot be reverted, and (\ref{mle_flat}) can be modified to
\begin{equation}
	\hat{k} = \arg\max_{k \, \in \, \mathbb{K}} \big| R(k)  \big|.
\end{equation}

\subsubsection{Frequency Selective Channel}

Under frequency selective channel,  $f_h\left(x_k[n]\right) = h[n] * x_k[n]$, where $h[n]$ represents the discrete-time channel impulse response that contains $L$ complex-valued taps.
Assuming that channel estimation is available at the receiver, the estimated data symbols can be obtained from (\ref{mle}) by implementing a search over $\mathbb{K}$.
However, in order to derive a low complexity solution from (\ref{mle}), one needs to assume that a \ac{CP} is appended to the beginning of the transmit signal, and $N_{\mathrm{CP}} > L-1$, where $N_{\mathrm{CP}}$ is the length of the \ac{CP} in samples.
Here, a vector notation is adopted for the sake of simplicity.
Let $\mathbf{H} \in \mathbb{C}^{N\times N}$ be the circulant channel matrix obtained from $h[n]$, $\mathbf{y}\in \mathbb{C}^{N\times 1}$ and $\mathbf{c} \in \mathbb{C}^{N\times 1}$ represent vectors whose entries are the samples from $y[n]$ and $c[n]$, respectively.
Thus, the estimated data symbol can be obtained by
\begin{equation} \label{mle_selective}
	\hat{k} = \arg\max_{k \, \in \, \mathbb{K}} \, \Re\left\lbrace \tilde{\mathbf{r}}(k) \right\rbrace,
\end{equation}
where $\tilde{\mathbf{r}}(k) = \mathcal{F}\left\lbrace \mathbf{c}^{\mathrm{H}}\mathbf{H}^{\mathrm{H}}\mathbf{y} \right\rbrace$, and $(\cdot)^{\mathrm{H}}$ is the Hermitian operator. Note that the channel matrix can be estimated at the receiver by transmitting known sequences.
\section{CSS-based Modulation Schemes} \label{sec:alternative}

\subsection{\ac{LoRa} Physical Layer} \label{subsec:lora}

\begin{figure*}[t!]
	\centering
	\includegraphics[]{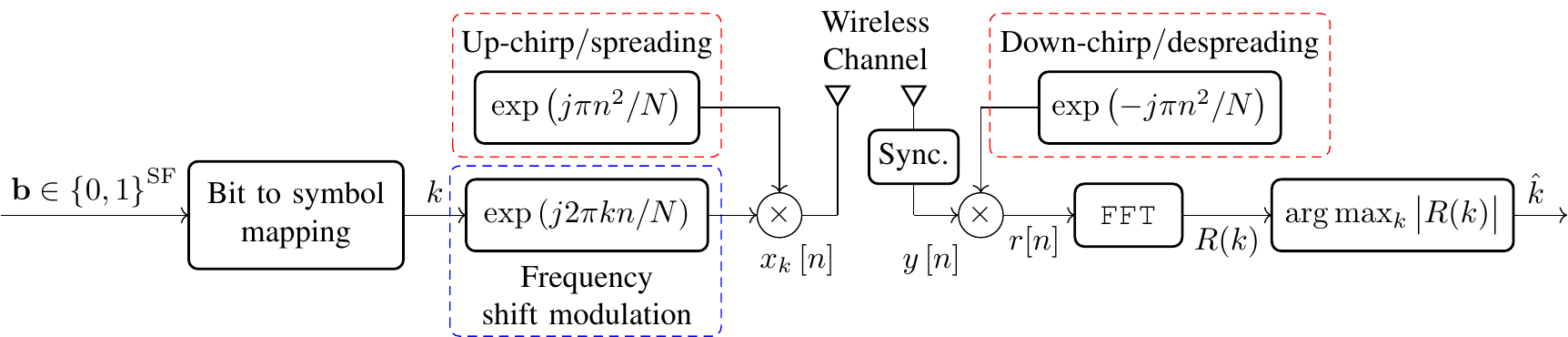}
	\caption{\ac{LoRa} \ac{PHY} transceiver block diagram.}
	\label{LoRATXRX}
\end{figure*}

The use of \ac{CSS} in the \ac{LoRa} wide area network standard is a recent and successful example of employment of chirp signals to wireless communications.
Notably, \ac{LoRa}'s \ac{PHY} employs \ac{CSS} in conjunction with a variant of \ac{FSK} modulation as described above \cite{loradatasheet,the_lora_modulation}. 
It is important to note that in this case, the discrete chirp rate is set to the unity, and the receiver employs non-coherent detection.
Moreover, \ac{LoRa} \ac{PHY} defines the \ac{SF} as the amount of bits that one symbol carriers, which ranges from 6 to 12 bits. 
In short, Fig. \ref{LoRATXRX} presents the discrete-time baseband \ac{LoRa} \ac{PHY} transceiver block diagram. 
The bit-word $\mathbf{b}$ contains SF bits that are mapped into one symbol $k$, which feeds the \ac{CSS} modulator. 
The despreading operation at the receiver side is accomplished by multiplying the received signal with a down-chirp, which is obtained by conjugating the up-chirp signal. 
At the receiver side, the estimated data symbol is obtained by selecting the frequency index with maximum value. 

The spreading gain, also known as processing gain, is defined by the ratio between the bandwidths of the spreading signal and  of the information signal, and for \ac{LoRa}'s \ac{PHY} it can be defined in dB as 
\begin{equation} \label{spreading_gain}
	G = 10\log_{10}\left(\frac{N}{\mathrm{SF}}\right),
\end{equation}
and as $N = 2^{\mathrm{SF}}$, the \ac{SF} also directly relates to the performance improvement observed with a spread spectrum system when compared to a non-spread system.
This is analogous to the performance enhancement achieved via coding schemes.
Larger \acp{SF} correspond to lower code rates, and the improvement obtained comes at the cost of reduced data rate.
Therefore, the \ac{SF} is used as adaptive modulation parameter depending on the \ac{SNR}, and consequently larger \acp{SF} allow longer coverage ranges with reduced data rates.

A key aspect of \ac{LoRa} \ac{PHY} is the fact that channel estimation and equalization are not necessary, since it employs the non-coherent detection receiver. 
However, employing coherent detection will improve the \ac{EE} performance of \ac{LoRa}, since the imaginary noise component is not taken in the estimation of the received data symbol.
Furthermore, under harsher multi-tap, i.e., frequency selective, channel conditions the original \ac{LoRa} modulation performance can degrade significantly, as also observed in \cite{facts_lora}.

\subsection{In-phase and Quadrature \ac{CSS}} \label{subsec:iqcss}

In \ac{IQCSS},  information is encoded in both in-phase (real) and quadrature (imaginary) components of the transmit signal \cite{iqcss}.
By making use of the orthogonality between the sine and cosine waves it is possible to transmit simultaneously two data symbols.
Its transmit signal is given by
\begin{equation} \label{tx_signal_IQCSS}
	x_{k_i, k_q}[n] = \sqrt{\frac{E_s}{2N}} g_{k_i, k_q}[n] c[n],
\end{equation}
where 
\begin{equation} \label{index_signal_IQCSS}
	g_{k_i, k_q}[n] = \exp \left( j\frac{2\pi }{N}k_in \right) + j\exp \left( j\frac{2\pi }{N}k_qn \right),
\end{equation}
where $k_i$ and $k_q$ are independent identically (uniform) distributed data symbols drawn from $\mathbb{K}$, and each carries $\mathrm{SF}$ bits. 
Thus, the total amount of transmitted bits is doubled when compared to the \ac{LoRa} \ac{PHY} specification.

\begin{figure*}[ht!]
	\centering
	\includegraphics[]{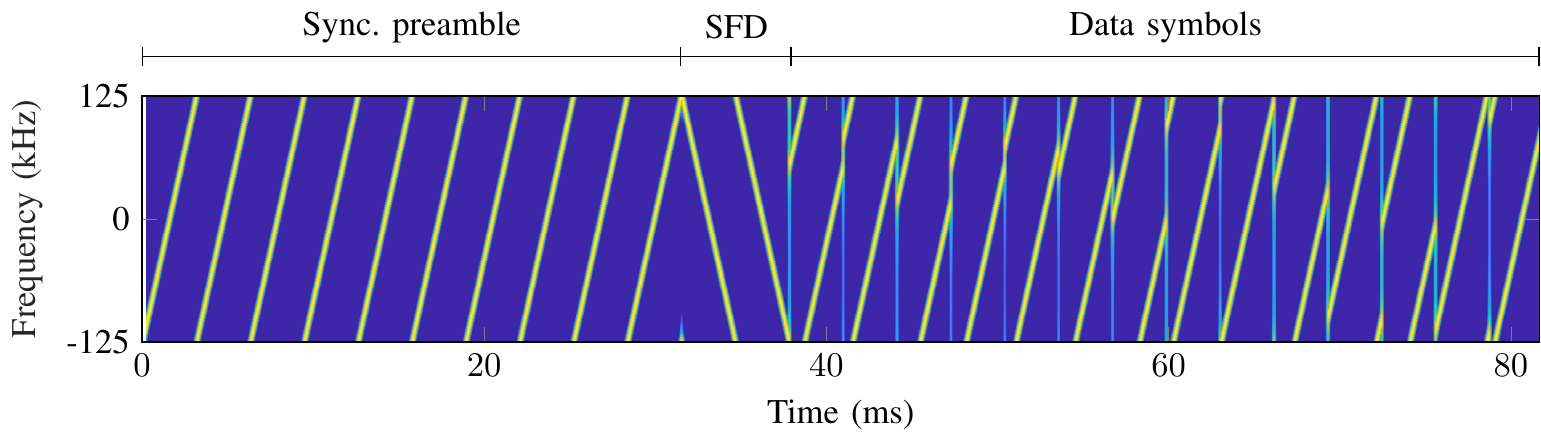}
	\caption{Spectrogram of the \ac{LoRa} \ac{PHY} packet structure.}
	\label{LoRa_packet}
\end{figure*}

\begin{figure*}[t!]
	\centering
	\includegraphics[]{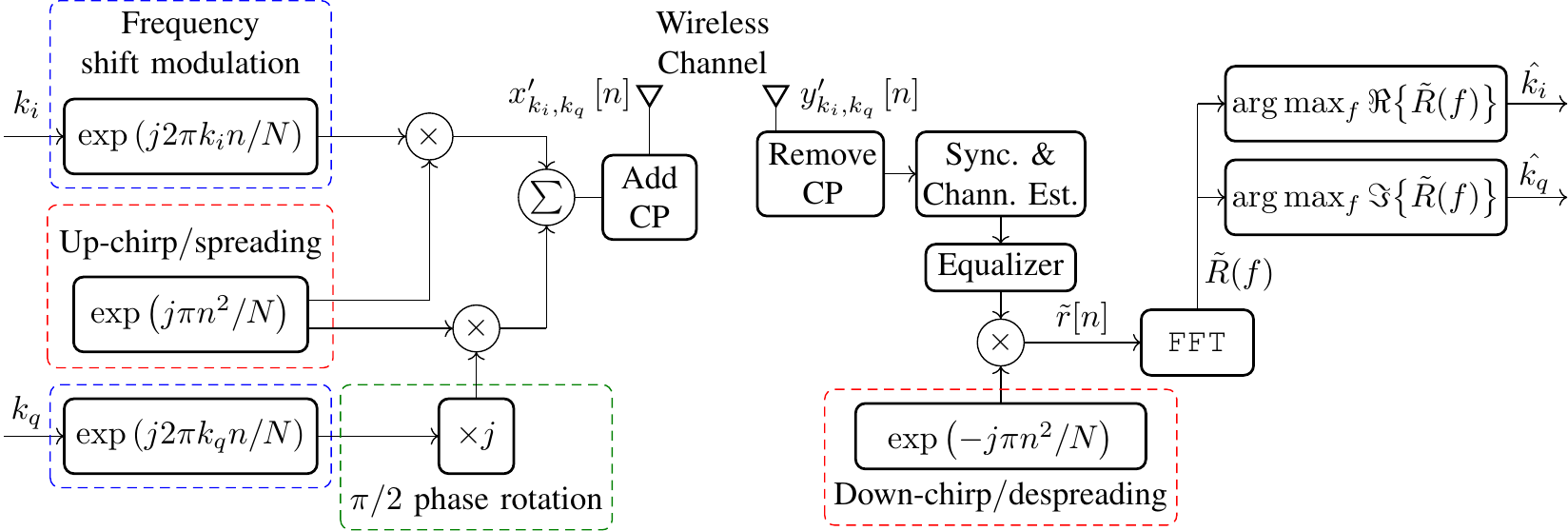}
	\caption{\ac{IQCSS} transceiver block diagram.}
	\label{IQCSS_TXRX}
\end{figure*}

Fig. \ref{IQCSS_TXRX} illustrates the discrete-time baseband \ac{IQCSS} transceiver block diagram. 
Note that the additional operations performed at the receiver side do not require modifications on the \ac{LoRa}'s transmit signal structure,  since \ac{IQCSS} makes use of the already available synchronization preamble for channel estimation and subsequent coherent detection. 
Fig. \ref{LoRa_packet} illustrates one \ac{LoRa} \ac{PHY} packet, where there are 14 modulated chirps in the data payload, 10 up-chirps are available for synchronization, followed by 2 down-chirps that indicate the beginning of the data symbols, and are named \ac{SFD} \cite{loradatasheet}.

The received signal after equalization and despreading is given by
\begin{equation}
	\tilde{r}[n] = g_{k_i, k_q}[n] + \tilde{w}[n],
\end{equation}
where 
and the received data symbols are given by
\begin{equation}
	\hat{k_i} = \arg\max_{f \in \mathbb{K}} \Re\big\lbrace \tilde{R}(f)\big\rbrace,
\end{equation}
\begin{equation}
	\hat{k_q} = \arg\max_{f \in \mathbb{K}} \Im\big\lbrace \tilde{R}(f)\big\rbrace,
\end{equation}
where $\tilde{R}(k) = \mathcal{F}\left\lbrace \tilde{r}[n]\right\rbrace$.

It is important to point out that \ac{IQCSS} requires coherent detection to work.
Therefore, for making further use of the information carried with the synchronization preamble, we propose to use the \ac{LS} approach for estimating the channel gain using the already available preamble structure for synchronization. 

Assuming that the channel presents flat-fading within its bandwidth, the received preamble can be written as
\begin{equation}
	y_p[n] = hx_p[n] + w[n],
\end{equation}
where $x_p[n]$ represents the 10 up-chirps transmitted at the beginning of the \ac{LoRa} \ac{PHY} packet, $w[n]$ is \ac{AWGN} with zero mean and $\sigma^2_w$ variance, and $h$ is the complex-valued channel gain.
Consequently, the \ac{LS} error criterion, which is the squared difference between the received data and the signal model \cite{kay1993fundamentals}, is given by
\begin{equation} \label{ls_error}
	J(h) = \sum_{n=0}^{N_p-1} \left( y_p[n] - hx_p[n] \right)^2,
\end{equation}
where $N_p=10N$ is the preamble length in samples.
Differentiating (\ref{ls_error}) w.r.t. $h$, and setting the result to zero yields to 
\begin{equation} \label{channel_estimator}
	\hat{h} = \frac{\mathbf{x}_p^{\mathrm{H}}\mathbf{y}_p}{\mathbf{x}_p^{\mathrm{H}}\mathbf{x}_p},
\end{equation}
where $\mathbf{x}_p$ and $\mathbf{y}_p$ are $N_p\!\times\!1$ vectors whose entries are the samples from $x_p[n]$ and $y_p[n]$, respectively, and $\hat{h}$ is the estimated channel gain.

For the case of frequency selective channels, the received preamble is given by
\begin{equation} \label{received_preamble_freq_selec}
	y_p[n] = h[n]*x_p[n] + w[n],
\end{equation}
where $h[n]$ represents the discrete-time channel impulse response with $L$ taps.
Under the assumption of a CP appended to the beginning of each chirp, the linear convolution in \eqref{received_preamble_freq_selec} becomes circular.
Thus, after CP removal, the $i$-th received chirp from the preamble can be written in matrix form as
\begin{equation} \label{received_preamble_freq_selec_matrix}
	\mathbf{y}_p^{(i)} = \mathbf{H}\mathbf{x}_p^{(i)} + \mathbf{w},
\end{equation}
where $\mathbf{x}_p^{(i)} \in \mathbb{C}^{N\times 1}$ contains the samples from the $i$-th up-chirp in the synchronization preamble.
Moreover, considering the commutative property of the convolution, \eqref{received_preamble_freq_selec_matrix} can be reformulated as
\begin{equation}
	\mathbf{y}_p^{(i)} = \mathbf{C}\mathbf{h} + \mathbf{w},
\end{equation}
where $\mathbf{C} \in \mathbb{C}^{N\times N}$ is a circulant matrix obtained from one raw up-chirp, and $\mathbf{h} \in \mathbb{C}^{N\times 1}$ contains the channel impulse response in its first $L$ entries.
Therefore, the \ac{LS} approach can be extended to frequency selective channels.
However, as mentioned, a \ac{CP} needs to be added to the preamble chirps, thus modifying the original \ac{LoRa} \ac{PHY} signal structure. 
Similarly to \eqref{ls_error}, the \ac{LS} error criterion can be written as
\begin{equation} \label{ls_error_vector}
	J(\mathbf{h}) = (\mathbf{y}_p^{(i)} - \mathbf{C}\mathbf{h})^{\mathrm{H}}(\mathbf{y}_p^{(i)} - \mathbf{C}\mathbf{h}).
\end{equation}
Differentiating \eqref{ls_error_vector} w.r.t. $\mathbf{h}$, setting the result to zero, and solving for $\mathbf{h}$ yields the \ac{LS} estimator of the channel impulse response as
\begin{align} \label{channel_estimator_selective_inv}
	\hat{\mathbf{h}} = \left( \mathbf{C}^{\mathrm{H}} \mathbf{C} \right)^{-1}\mathbf{C}^{\mathrm{H}} \mathbf{\bar{y}}_p,
\end{align}
where $\hat{\mathbf{h}}$ contains the estimated channel impulse response, and we define 
\begin{equation}
	\mathbf{\bar{y}}_p = \frac{1}{P}\sum_{i=1}^{P} \mathbf{y}_p^{(i)}
\end{equation}
as a $N\!\times\!1$ vector whose entries are the averaged samples over the $P$ received synchronization chirps after CP removal. 
This allows noise reduction, since all chirps in the preamble are used for estimating the channel.
Note that $P=10$ up-chirps following the \ac{LoRa} \ac{PHY} structure.
It is important to note that, $\hat{\mathbf{h}}$ has in total $N$ entries, but only the first $L$ are used as an estimate of the channel taps.
Furthermore, by inspecting (\ref{channel_estimator_selective_inv}), one can see that $\mathbf{C}$ is an orthogonal matrix, and the estimator simplifies to
\begin{align} \label{channel_estimator_selective}
	\hat{\mathbf{h}} = \mathbf{C}^{\mathrm{H}} \mathbf{\bar{y}}_p,
\end{align}
which reduces considerably the estimation computational complexity. 

Assuming that the channel estimation and equalization modules are available at the receiver, coherent detection can be made possible.
In this case, the receiver structure allows greater SE and EE when compared with \ac{LoRa} modulation. 
Nevertheless, if these modules are not used, the receiver structure is equivalent to the \ac{LoRa} standard.
Therefore, the receiving devices (gateways) that operate using \ac{IQCSS} can still decode the information transmitted using the original \ac{LoRa} modulation, thus making \ac{IQCSS} gateways backwards compatible. 

\subsection{Discrete Chirp Rate Keying \ac{CSS}} \label{subsec:dcrk}

While \ac{IQCSS} is able to increase the data throughput by exploiting the benefits of a coherent detection receiver, it is also possible to encode extra information bits on the discrete chirp rate and maintain simpler non-coherent detection.
A similar approach is has been proposed in \cite{ssk}. 
However, this approach is limited to only one extra information bit.
Hence, we propose to extend this framework to multiple extra information bits.
Hereafter, we refer to this scheme as \ac{DCRK-CSS}.

Let $\mathbf{b}_e \in \left\lbrace 0,\,1 \right\rbrace^{N_e}$ denote the bit-word that will modulate the chirp rate, where $N_e$ represents the amount of extra bits encoded, and $\mathbf{b}_f \in \left\lbrace 0,\,1 \right\rbrace^{N_f}$ denote the bit-word that will modulate the frequency of a complex exponential, $N_f$ is equivalent to \ac{LoRa}'s spreading factor.
The total amount of possible chirp rates is given by $P = 2^{N_e}$.
Therefore, the data symbol is comprised by both bit words, i.e., $\mathbf{b} = \left\lbrace \mathbf{b}_f,\, \mathbf{b}_e \right\rbrace^{N_f+N_e}$.

The discrete chirp rate is selected depending upon the combination a of bits in $\mathbf{b}_e$, e.g., for $N_e = 3$ bits, the defined chirp rates are given by
\begin{equation}
	M_p = \left\{ \begin{array}{rcl}
		-4 & \mbox{for} & \mathbf{b}_e = \left[0\,0\,0\right] \\ 
		-3 & \mbox{for} & \mathbf{b}_e = \left[0\,0\,1\right] \\
		-2 & \mbox{for} & \mathbf{b}_e = \left[0\,1\,0\right] \\
		-1 & \mbox{for} & \mathbf{b}_e = \left[0\,1\,1\right] \\
		1 & \mbox{for} & \mathbf{b}_e = \left[1\,0\,0\right] \\ 
		2 & \mbox{for} & \mathbf{b}_e = \left[1\,0\,1\right] \\
		3 & \mbox{for} & \mathbf{b}_e = \left[1\,1\,0\right] \\
		4 & \mbox{for} & \mathbf{b}_e = \left[1\,1\,1\right],
	\end{array}\right.
\end{equation}
where $M$ assumes $P$ non-zero integer values, and $p$ is an indexing variable.
The transmit signal can be written as
\begin{equation}
	x_{k, M}[n] =\sqrt{\frac{E_s}{N}} \exp \left( j\frac{2\pi }{N}kn \right) c_p[n],
\end{equation}
where $c_p[n] = \exp\left( j\pi M_p n^2/N \right)$.

\begin{figure*}[t!]
	\centering
	\includegraphics[]{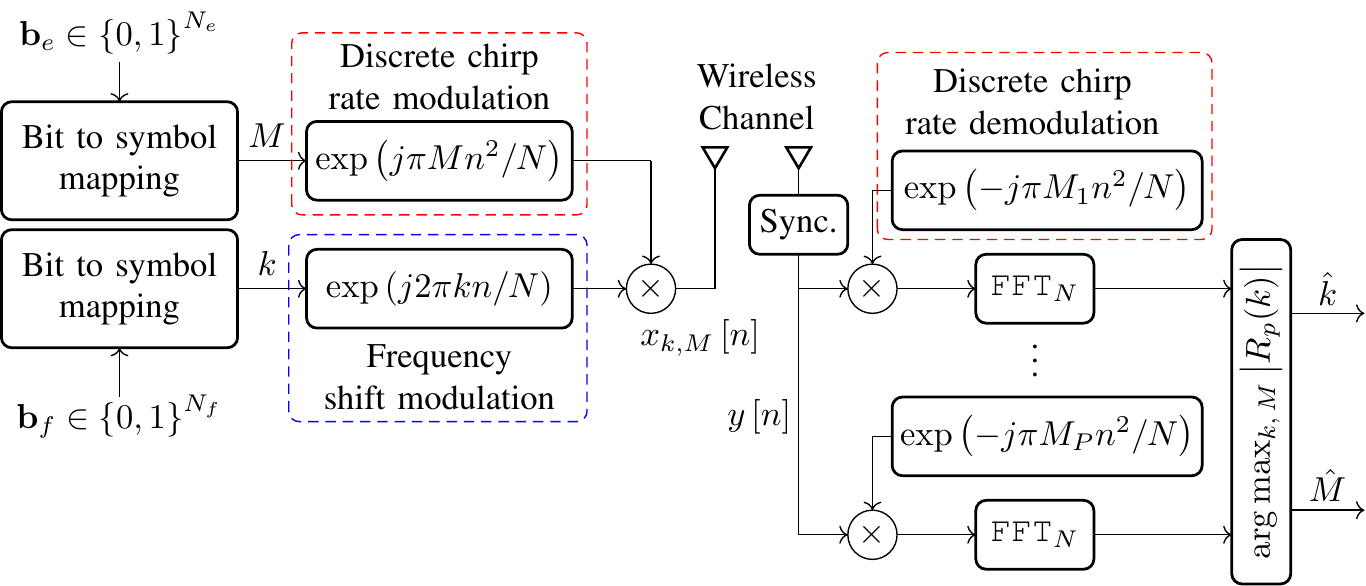}
	\caption{Discrete Chirp Rate Keying \ac{CSS} transceiver block diagram assuming non-coherent detection.}
	\label{DCRKCSS}
\end{figure*}

Fig. \ref{DCRKCSS} illustrates the discrete-time baseband transceiver block diagram.
The received signal after synchronization goes through a bank of dechirping modules with different chirp rates.
Let the \acp{DFT} of despreaded signals be represented by
\begin{equation}
	R_p(k) = \mathcal{F}\left\lbrace y[n] \exp\left(- j\pi M_p n^2/N \right) \right\rbrace, \; \mathrm{for} \; 1\le p\le P,
\end{equation}
where $y[n] = hx_{k, M}[n] + w[n]$ considering transmission over a flat fading channel.
The estimated frequency, and chirp encoded data symbols are jointly obtained as
\begin{equation}
	\hat{k}, \; \hat{M} = \arg\max_{k,\, M} \big|R_p(k)\big|.
\end{equation}

\begin{figure*}[t!]
	\centering
	\includegraphics[]{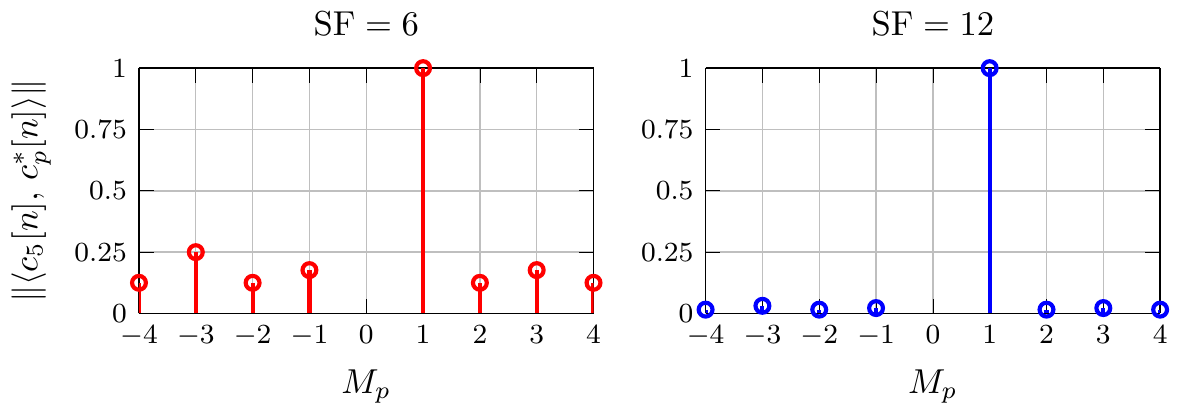}
	\caption{Inner product between $c_5[n]$ and different dechirping rates, considering $\mathrm{SF}=6$ and $\mathrm{SF}=12$.}
	\label{inner_prod}
\end{figure*}
It is important to note that the spreading chirps with different rates are not orthogonal, but they present low correlation, and the performance loss due to intrinsic interference becomes negligible.
Furthermore, the level of intrinsic interference is inversely proportional to $N$, such that smaller \acp{SF} are subject to more performance loss.
To better illustrate this characteristic, Fig. \ref{inner_prod} shows the absolute value of the inner product between $c_5[n]$ and the despreading signals defined by
\begin{equation}
	\langle c_5[n],\, c_p^*[n] \rangle = \sum_{n=0}^{N-1} \exp\left( j\pi n^2/N \right) \exp\left(- j\pi M_p n^2/N \right).
\end{equation}

\section{Performance Analysis} \label{sec:performance}

For comparing the performance of the proposed schemes against the \ac{LoRa} \ac{PHY}, we resort to numerical simulations for estimating the symbol bit error ratio (SER/BER) under three different wireless channel models, namely \ac{AWGN}, time-variant (TV) non-frequency-selective (Rayleigh) channel, and time-variant frequency-selective (TVFS). 
For the latter, the channel model chosen is "Typical case for urban area" with 12 taps \cite{gsm_channel}. 
This has been chosen due to the similar operating frequency and bandwidth of \ac{LoRa} and \ac{GSM} in Europe.
Moreover, based on the symbol error ratio, the maximum achievable throughput is also presented for supporting the claims of increased spectral efficiency. 

Table \ref{simulation_parameters} shows the simulation parameters considered.
\begin{table}[h]
	\centering
	\caption{Simulation parameters}
	\label{simulation_parameters}
	\renewcommand{\arraystretch}{1.5}
	\begin{tabularx}{\columnwidth}{ll}
		\toprule[0.9pt] 
		Parameter & Value \\ 
		\midrule
		Spreading factor & $\mathrm{SF} \in \left\lbrace 6, \; 9, \; 12\right\rbrace$ \\
		Extra information bits & $N_e \in \left\lbrace 2,\; 3 \right\rbrace $ \\
		Bandwidth & 250 kHz \\ 
		Carrier frequency & 863 MHz \\ 	
		Mobile speed & 3 km/h \\	
		CP length & 16 samples \\ 
		TV channel & single tap (Rayleigh) \\
		TVFS channel & 12 taps "Typical case for urban area" \cite{gsm_channel} \\
		Data frame size & 30 chirps \\
		Preamble frame size & 10 up chirps + 2 down chirps \\
		\bottomrule[0.9pt]	
	\end{tabularx}
\end{table}

\subsection{Symbol and Bit Error Ratio Analysis}

\subsubsection{AWGN Channel}

\begin{figure*}[t!]
	\centering
	\begin{subfigure}[b]{8.7cm}
		\centering
		\includegraphics[]{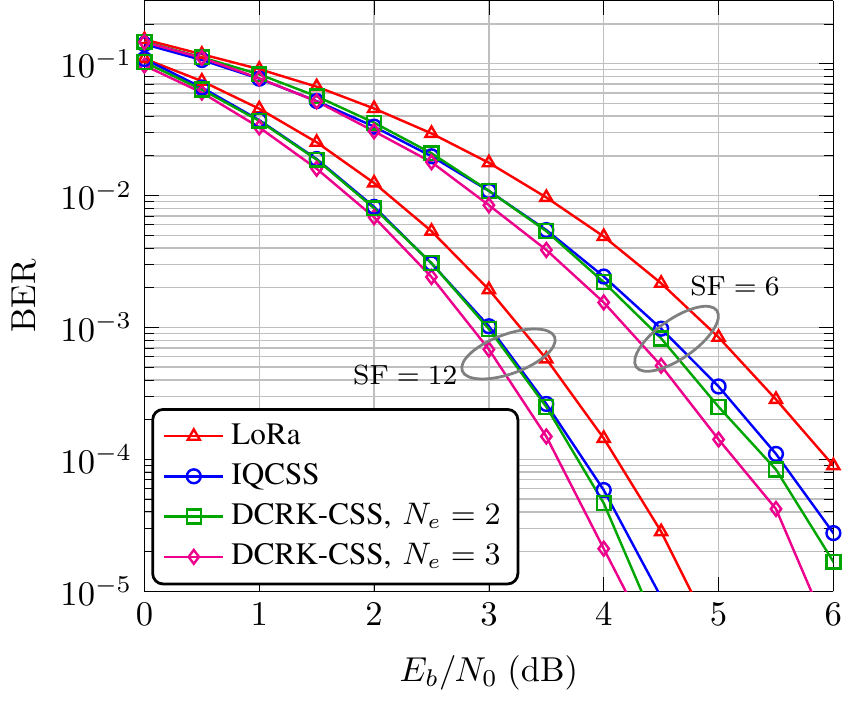}
		\caption{Average bit energy versus bit error ratio.}
		\label{BERvsEbN0}
	\end{subfigure}
	\hfill
	\begin{subfigure}[b]{8.7cm}
		\centering
		\includegraphics[]{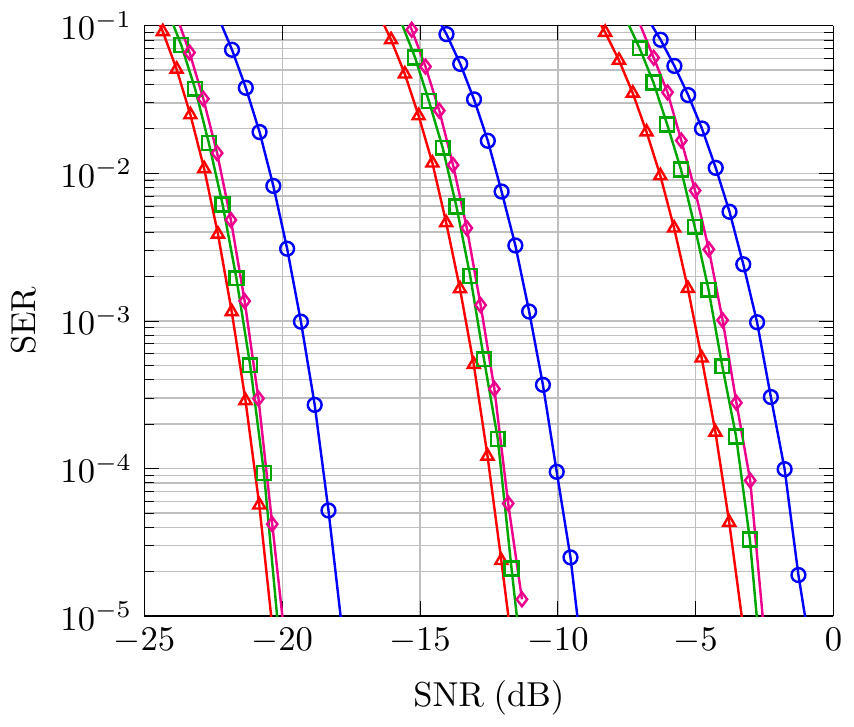}
		\caption{Signal-to-noise ratio versus symbol error ratio.}
		\label{SERvsSNR}
	\end{subfigure}
	\caption{Error performance under \ac{AWGN} channel.}
\end{figure*}

Fig. \ref{BERvsEbN0} shows the estimated BER under \ac{AWGN} channel for \ac{LoRa}, \ac{IQCSS} and \ac{DCRK-CSS}. 
Note that \ac{IQCSS} transmits twice the amount of bits when compared with \ac{LoRa}, and \ac{DCRK-CSS} transmits $N_e$ extra bits.  
Larger \acp{SF} will result in better performance, since all schemes transmit data using frequency modulation, and adding more symbols to the constellation does not reduce the minimum symbol distance.
There is a gap of about 0.5 dB between the curves of \ac{IQCSS} and \ac{LoRa} for the same \ac{SF}, whereas between \ac{LoRa} and \ac{DCRK-CSS} the gap increases with $E_b/N_0$.
The gap between \ac{LoRa} and \ac{IQCSS} is observed because \ac{IQCSS} collects less noise than \ac{LoRa} in the process of detection, since it explores the phase information instead of making a decision based solely on the estimated energy of the frequency bins.
Thus, one can use less energy to transmit more information with the same bit error probability when employing \ac{IQCSS} over \ac{LoRa}. 
However, the gap between \ac{LoRa} and \ac{DCRK-CSS} is observed due to the smaller average energy bit required to maintain an equal BER, and this gap increases as the number of extra bits encoded on the chirp slope increases. 
Nevertheless, note that for the lower $E_b/N_0$ regime, the gap is reduced when compared with the high $E_b/N_0$ regime, this is observed since the different despreading signals are not completely orthogonal to each other, but rather present a significant low correlation.

Fig. \ref{SERvsSNR} also shows SER under \ac{AWGN} channel while taking into consideration the spreading gain.
The transmit power available is divided between the in-phase and quadrature components for \ac{IQCSS}, while in \ac{LoRa} and \ac{DCRK-CSS} all power is used for a single component.
Therefore, there is a 3 dB gap between \ac{LoRa} and \ac{IQCSS} in Fig. \ref{SERvsSNR}. 
Moreover, there is small gap between \ac{LoRa} and \ac{DCRK-CSS}, which is present due to the intrinsic interference caused by the non-orthogonality between the different despreading signals.
As we can see, the performance degradation is acceptable for the increased throughput provided by \ac{DCRK-CSS}.
It is also important to point out that for larger \acp{SF}, the this gap is reduced, since the intrinsic interference is reduced as the \ac{SF} increases.

\begin{figure}[t!]
	\centering
	\includegraphics[]{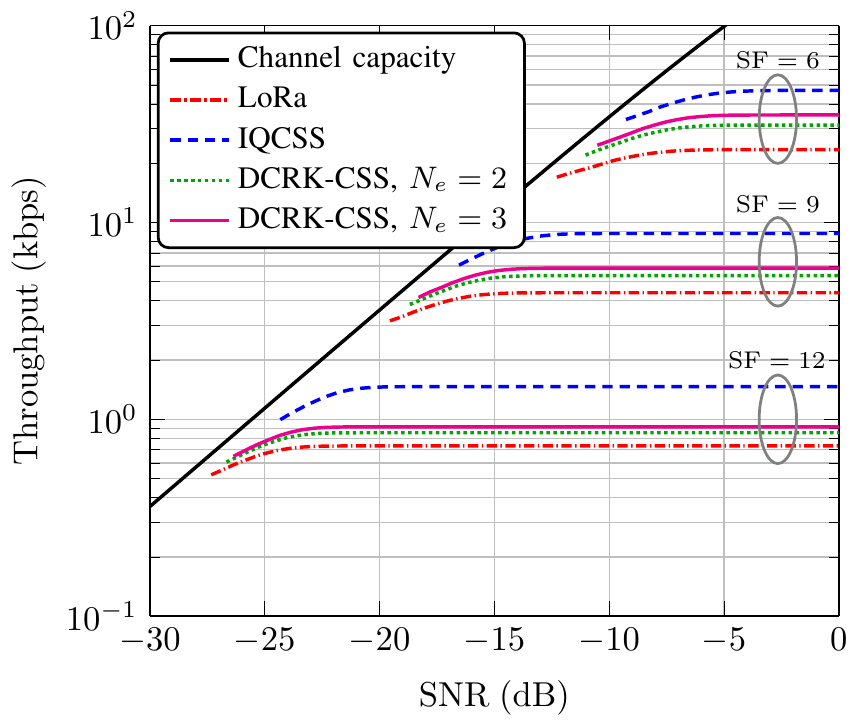}
	\caption{Maximum throughput achievable under AWGN channel.}
	\label{capacity}
\end{figure}

In order to quantize the advantages with respect to the maximum throughput achievable by the presented schemes, Fig. \ref{capacity} presents the maximum achievable throughput under AWGN channel.
The Shannon channel capacity is also plotted for reference.
\ac{IQCSS} achieves more than double the maximum throughput of \ac{LoRa}, since it encodes double the amount of bits, and it benefits from coherent detection at the receiver. 
\ac{DCRK-CSS} is also able to increase the maximum throughput when compared to the \ac{LoRa} PHY scheme. 
The effective throughput (bps) is calculated as 
\begin{equation}
	\rho = \frac{N_b B}{N}(1 - \mathrm{BER}),
\end{equation}
where $N_b$ represents the number of bits carried by each modulation scheme.
Considering $\mathrm{SF}=12$, there is an increase on the throughput when comparing \ac{DCRK-CSS} with \ac{LoRa} of 16.6\% and 25\% for $N_e=2$ and $N_e=3$ bits, respectively.
Comparatively, for $\mathrm{SF}=6$ there is an increase of 33.3\% and 50\% for $N_e=2$ and $N_e=3$, respectively.
Therefore, both \ac{IQCSS} and \ac{DCRK-CSS} make a more efficient usage of spectrum resources.
These are key aspects required for more data demanding \ac{IoT} applications.

\begin{figure}[t!]
	\centering
	\includegraphics[]{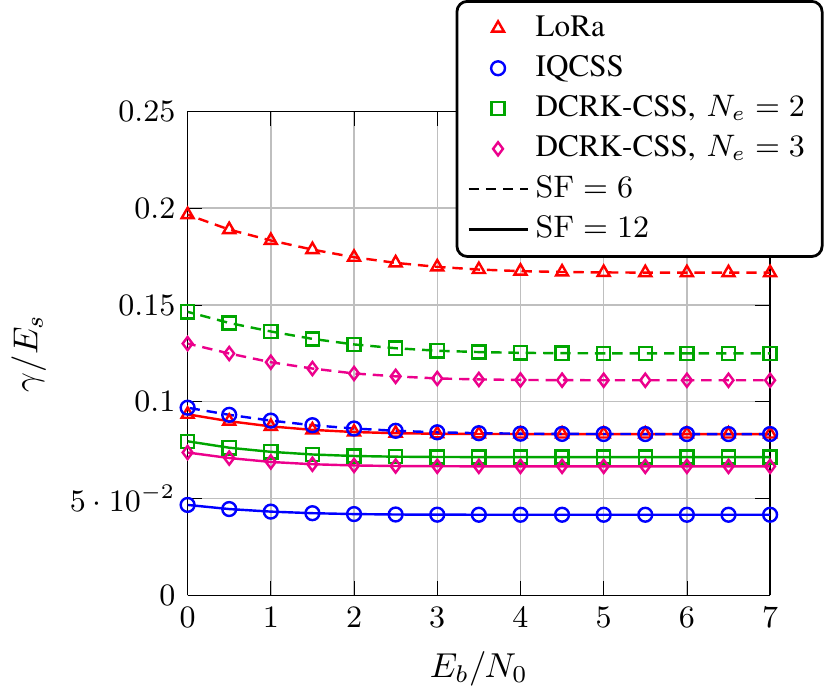}
	\caption{Effective energy per useful bits normalized to $E_s$ assuming transmission over AWGN channel.}
	\label{energy_efficiency}
\end{figure}

The energy efficiency of the modulation schemes can be indirectly demonstrated via Fig. \ref{BERvsEbN0}.
However, this figure does not show how much of the energy is used to transmit bits that are correctly received.
Therefore, Fig. \ref{energy_efficiency} presents the energy that is required to transmit with a certain bit error rate for the different modulation schemes.
The effective energy per useful bits is defined as
\begin{equation}
	\gamma = \frac{E_s}{N_b(1-\mathrm{BER})}.
\end{equation}
By inspecting the result in Fig. \ref{energy_efficiency}, smaller SFs are less energy efficient because they are less robust to bit errors, and as more bits are transmitted with the alternative schemes, the effective energy required to transmit with a certain bit error ratio is reduced when compared to the original \ac{LoRa}.

\subsubsection{Flat Fading Channel}

\begin{figure}[t!]
	\centering
	\includegraphics[]{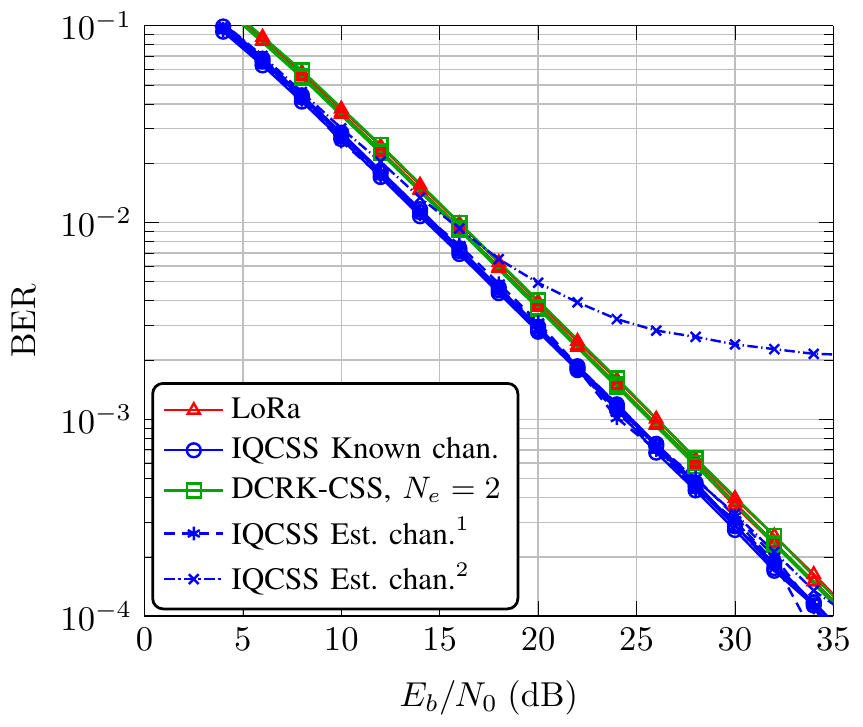}
	\caption{Average bit energy versus bit error ratio under flat fading channel.}
	\label{BERvsEbN0_TV}
\end{figure}

\begin{figure}[t!]
	\centering
	\includegraphics[]{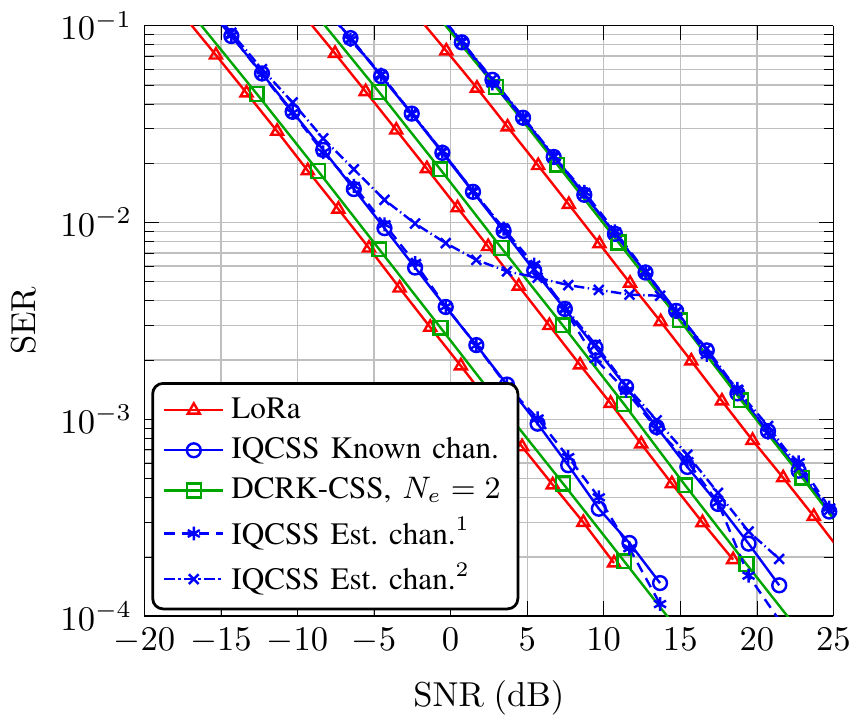}
	\caption{Signal-to-noise ratio versus symbol error ratio under flat fading channel.}
	\label{SERvsSNR_TV}
\end{figure}

Fig. \ref{BERvsEbN0_TV} and Fig. \ref{SERvsSNR_TV} show respectively the estimated BER and SER under time-variant flat fading channel. 
The same behavior observed under AWGN channel regarding the performance gaps can be observed between \ac{LoRa} and \ac{DCRK-CSS}.
Moreover, there are three BER and SER curves for accessing \ac{IQCSS}'s performance. 
The solid curves with circle marks represents the BER considering an unrealistic scenario where the channel coefficient is known at the receiver, i.e., this represents the baseline performance. 
Assuming that the channel coefficient remains static during the transmission of one frame, which has 10 chirps for synchronization and channel estimation, and 30 chirps encoded with information, the dashed curves with asterisk marks (Est. chan.$^1$) are obtained. 
These results show that the proposed channel estimation technique yields performance comparable to perfect channel estimation in cases where the channel complex gain does not change during the transmission of frame. 
Lastly, the dot dashed curves with x marks (Est. chan.$^2$) represent the case where the channel coefficient changes during the transmission of one frame. 
In this last case, we consider that the relative speed between transmitter and receiver is 3 km/h, carrier frequency is 863 MHz, the bandwidth occupied is 250 kHz.
As a result from the mobility, \ac{IQCSS} suffers a performance degradation when compared with \ac{LoRa} and \ac{DCRK-CSS}. 
However, considering a low mobility scenario, this degradation might be neglected. 
For this particular simulation scenario, significant performance degradation is observed for \acp{SF} greater than 10.

\subsubsection{Frequency Selective Channel}

\begin{figure}[t!]
	\centering
	\includegraphics[]{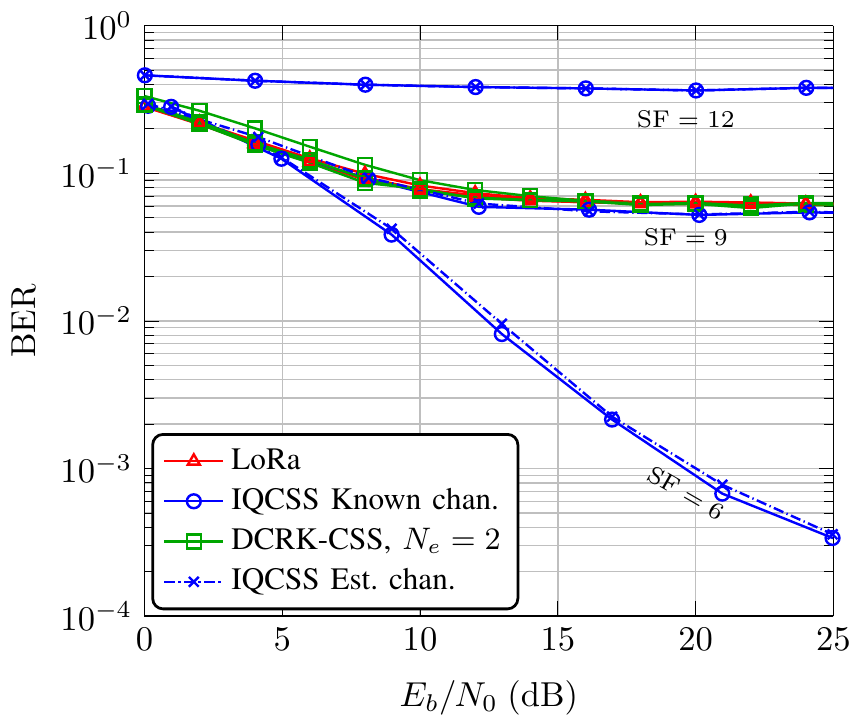}
	\caption{Average bit energy versus bit error ratio under frequency selective channel.}
	\label{BERvsEbN0_TVFS}
\end{figure}

Fig. \ref{BERvsEbN0_TVFS} shows the BER under the GSM channel model.
Due to multipath components of the channel, the \ac{DFT} output from \ac{LoRa}'s and \ac{DCRK-CSS}'s receivers will present multiple peaks, and depending on the channel realization, \ac{LoRa}'s receiver is not able to distinguish between them, but still presents a certain level of robustness.
As we can see, the performance of these two schemes degrades significantly under channels with long power delay profiles.
However, \ac{IQCSS} is able to deal with harsher multipath channels due to its more sophisticated receiver structure.
Two curves are presented for accessing \ac{IQCSS}'s performance under TVFS channels. 
The solid curve with circle marks is obtained considering noiseless channel estimation, whereas the dot dashed curve with x marks represents realistic case, and both use the LS channel estimation given by (\ref{channel_estimator_selective}).
As one can see, with LS channel estimation the BER performance approaches the case of ideal channel estimation.
An error floor is present, since the channel impulse response changes within the frame duration.
Hence, the smaller \acp{SF} present better performance due to its shorter time duration.
The mobility scenario is considered for all curves presented in Fig. \ref{BERvsEbN0_TVFS}.
For reference, a \ac{CP} with length $N_{\mathrm{CP}} = 16$ samples has been employed with \ac{IQCSS}.

From the performance analysis presented, one can conclude that employing \ac{IQCSS} in conditions where the wireless channel response remains static during the transmission of a data frame, and  employing \ac{DCRK-CSS} when it does not present significant frequency selectivity, would be advantageous for increasing the data throughput without sacrificing energy efficiency.
Hence, for increasing the spectral efficiency in conditions where high mobility of the transmitting units is more likely to occur when compared to frequency selectivity, \ac{DCRK-CSS} is a valid alternative to the \ac{LoRa} \ac{PHY} scheme.
Moreover, both receivers from \ac{IQCSS} and \ac{DCRK-CSS} are designed to be backwards compatible with the current \ac{LoRa} standard.

It is also important to point-out that the investigation shown above considers that all modulation schemes operate with the same transmission power.
For the sporadic transmission of a large amount of \ac{PHY} data packets, the transmitting module of the device will be operating for shorter periods of time, since \ac{IQCSS} and \ac{DCRK-CSS} can transmit more data per unit of time when compared to \ac{LoRa}.
Therefore, energy saving and longer battery life can be achieved with the proposed schemes.
However, an in-depth analysis of such gains is out of the scope of this paper.
\section{Conclusion} \label{sec:conclusions}

This paper has presented and studied novel modulation schemes inspired by the \ac{LoRa} \ac{PHY} for wireless applications that require low energy consumption and enhanced spectral efficiency. 
The major advantage of \ac{IQCSS} lies in the ability to double the \ac{SE}, and at the same time to improve \ac{EE} when compared with the conventional \ac{CSS} employed by \ac{LoRa} when employing coherent detection at the receiver side.
An estimation framework that uses the synchronization preamble defined by \ac{LoRa} \ac{PHY} to estimated the complex-valued channel gain, in flat fading channels, and an extended framework for frequency selective channels have also been presented.
On the condition of flat fading channels, the throughput can be enhanced by encoding extra information bits on the discrete chirp rate.
\ac{DCRK-CSS} is able to enhance \ac{SE} while maintaining the benefits of non-coherent detection.

A key aspect of \ac{IQCSS} and \ac{DCRK-CSS} is that their receivers are still able to decode the information transmitted using the original \ac{LoRa} modulation.
Hence, gateways designed to operate with either are backwards compatible with the conventional \ac{LoRa}.
Therefore, the investigations shown in this paper have demonstrated the potential of \ac{DCRK-CSS} and \ac{IQCSS} as solutions for enhancing the operation of \ac{LPWAN}.
Nevertheless, a study on the performance of the presented modulation techniques in conjunction with channel coding remains as an interesting topic to pursue.

\section*{Acknowledgment}
This research was supported by the European Union under the iNGENIOUS project, by DAAD, MESRI, and MEAE under the PROCOPE 2020 project, and by the CY Initiative through the ASIA Chair of Excellence under Grant PIA/ANR-16-IDEX-0008
\bibliographystyle{ieeetr}
\bibliography{my_references}
\vskip -2\baselineskip plus -1fil	
\begin{IEEEbiography}[{\includegraphics[width=1in,height=1.25in,clip,keepaspectratio]{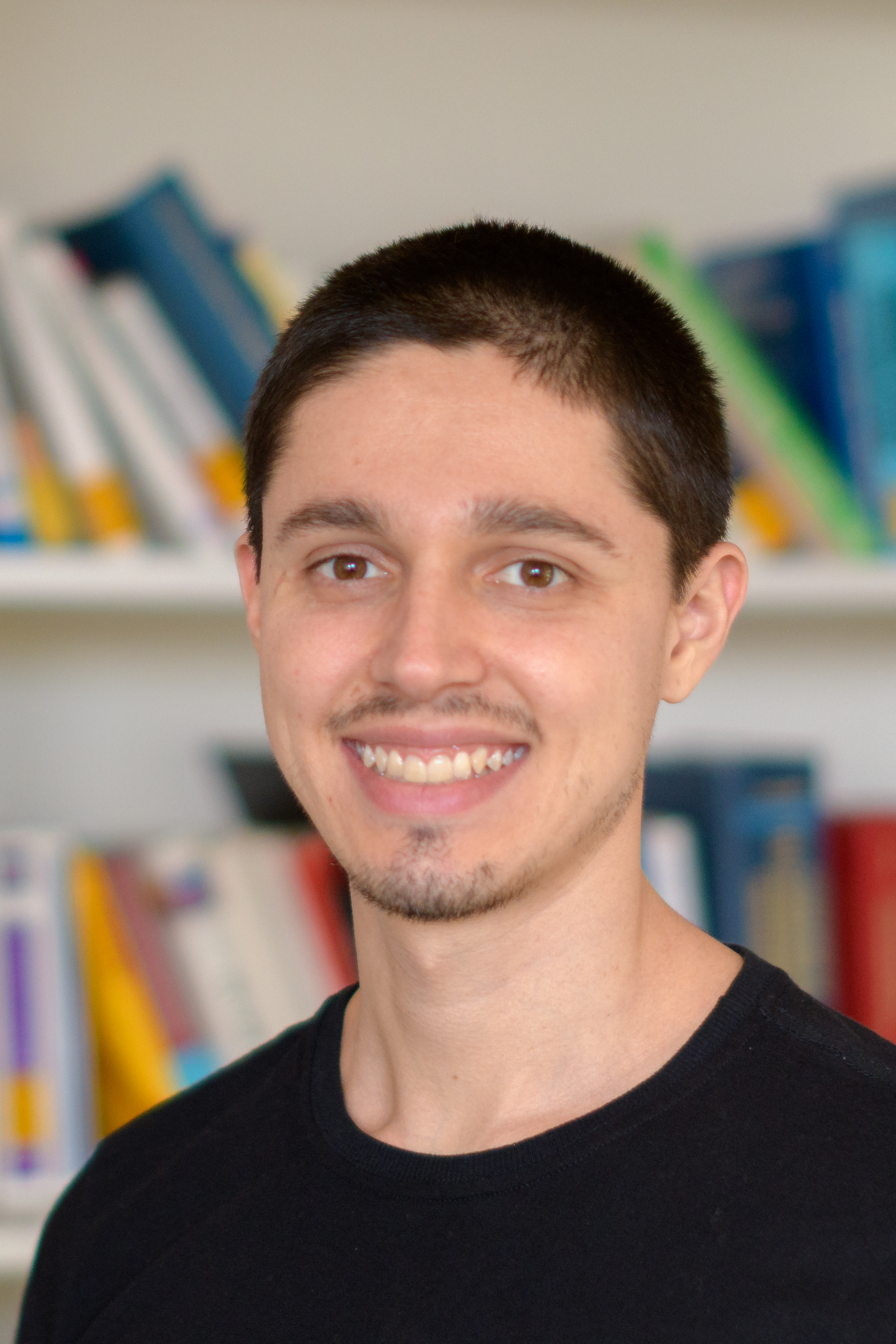}}]{Ivo Bizon Franco de Almeida}
	received the five-year BSc. and MSc. degrees in Electrical Engineering from the Instituto Nacional de Telecomunica\c{c}\~oes (Inatel), Brazil in 2016 and 2018, respectively. 
	Currently he is pursuing his PhD at Technische Universi\"at Dresden (TUD), Germany, and working as a research associate at the Vodafone Chair Mobile Communications Systems. His current research interests include deep learning based localization schemes and modulation techniques for future long range and low power wireless systems.
\end{IEEEbiography}
\vskip -2\baselineskip plus -1fil	
\begin{IEEEbiography}[{\includegraphics[width=1in,height=1.25in,clip,keepaspectratio]{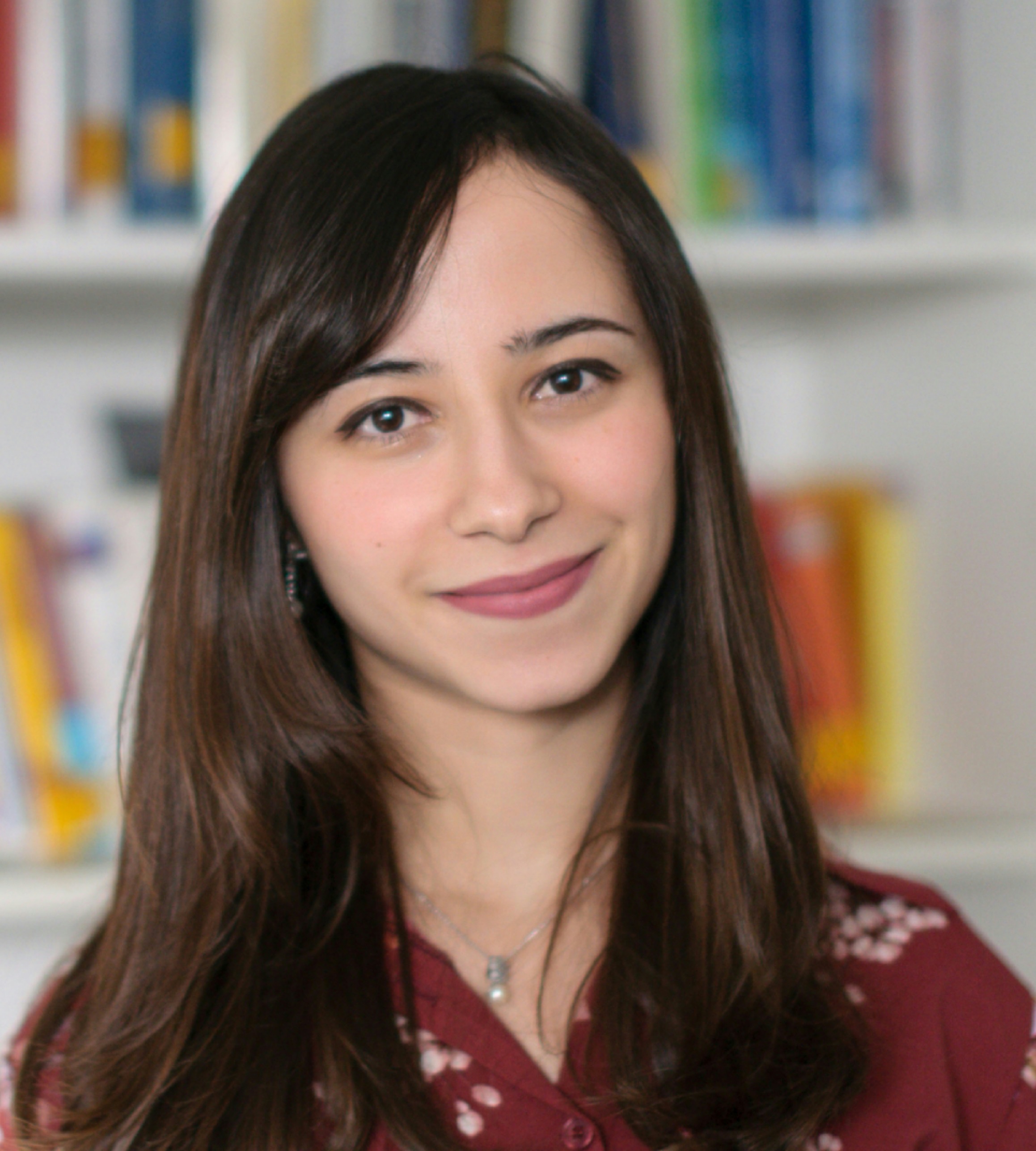}}]{Marwa Chafii}
	Marwa Chafii received her Ph.D. degree in electrical engineering in 2016, and her Master's degree in the field of advanced wireless communication systems (SAR) in 2013, both from CentraleSupélec, France. Between 2014 and 2016, she has been a visiting researcher at Poznan University of Technology (Poland), University of York (UK), Yokohama National University (Japan), and University of Oxford (UK). She joined the Vodafone Chair Mobile Communication Systems at the Technical University of Dresden, Germany, in February 2018 as a research group leader. Since September 2018, she is and an associate professor at ENSEA, France, where she holds a Chair of Excellence on Artificial Intelligence from CY Initiative. She received the prize of the best Ph.D. in France in the fields of Signal, Image \& Vision, and she has been nominated in the top 10 Rising Stars in Computer Networking and Communications by N2Women in 2020. Since 2019, she serves as Associate Editor at IEEE Communications Letters where she received the Best Editor Award in 2020. She is currently vice-chair of the IEEE ComSoc ETI on Machine Learning for Communications, leading the Education working group of the ETI on Integrated Sensing and Communications, and research lead at Women in AI.
\end{IEEEbiography}
\vskip -2\baselineskip plus -1fil	
\begin{IEEEbiography}[{\includegraphics[width=1in,height=1.25in,clip,keepaspectratio]{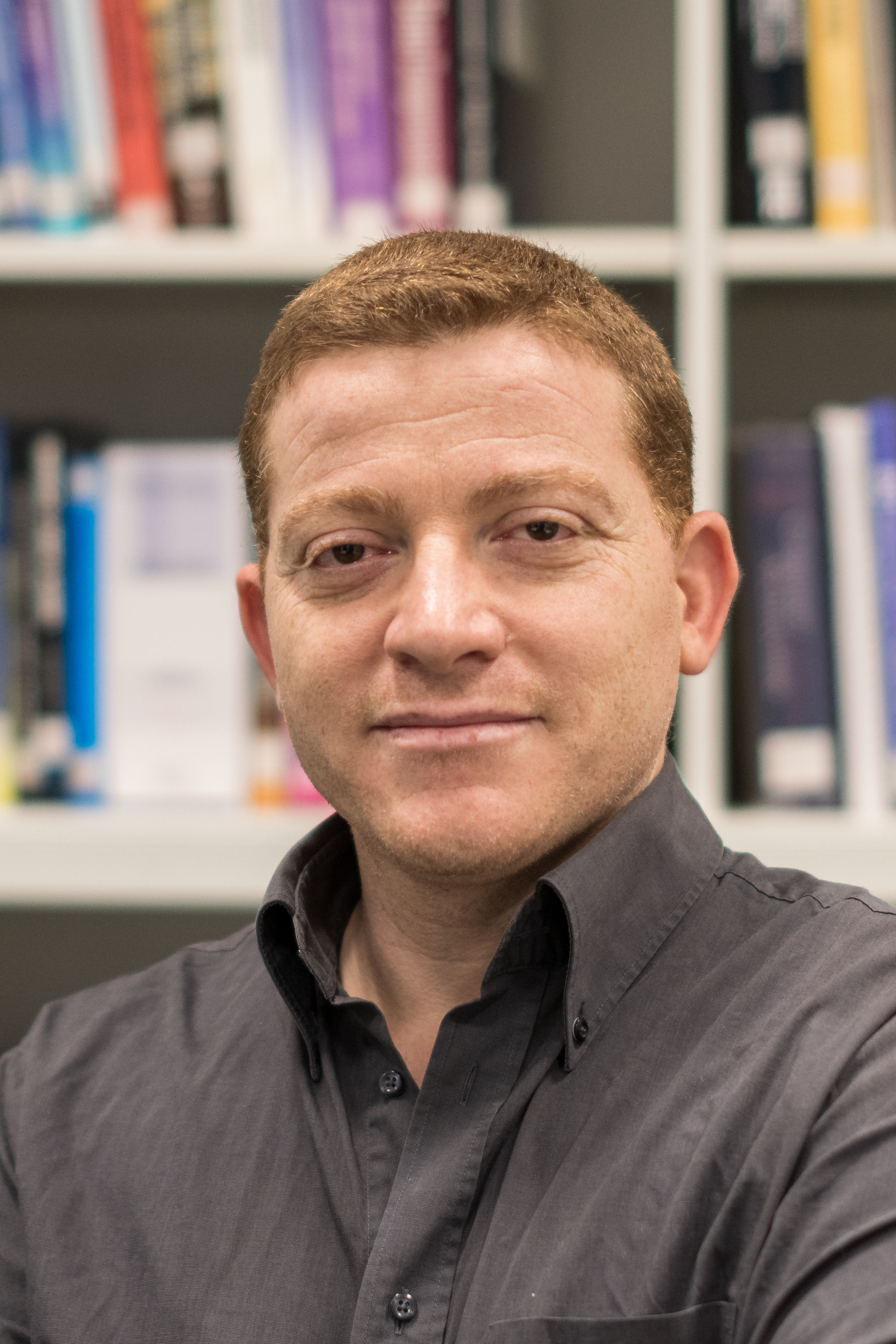}}]{Ahmad Nimr}
	Ahmad Nimr is a member of Vodafone Chair at TU Dresden since October 2015. His research activities focus on multicarrier waveforms and multiple access techniques. He is also involved in the design and implementation of real-time communication systems. Ahmad received his diploma in Communication Engineering in 2004. Afterwards, he worked as a software and hardware developer from 2005 to 2011. Then, he perused a master of science in Communications and Signal Processing and obtained his M.Sc degree in 2014 from TU Ilmenau. He received the best graduate student award for his excellent Master's grades.
\end{IEEEbiography}
\vskip -2\baselineskip plus -1fil	
\begin{IEEEbiography}[{\includegraphics[width=1in,height=1.25in,clip,keepaspectratio]{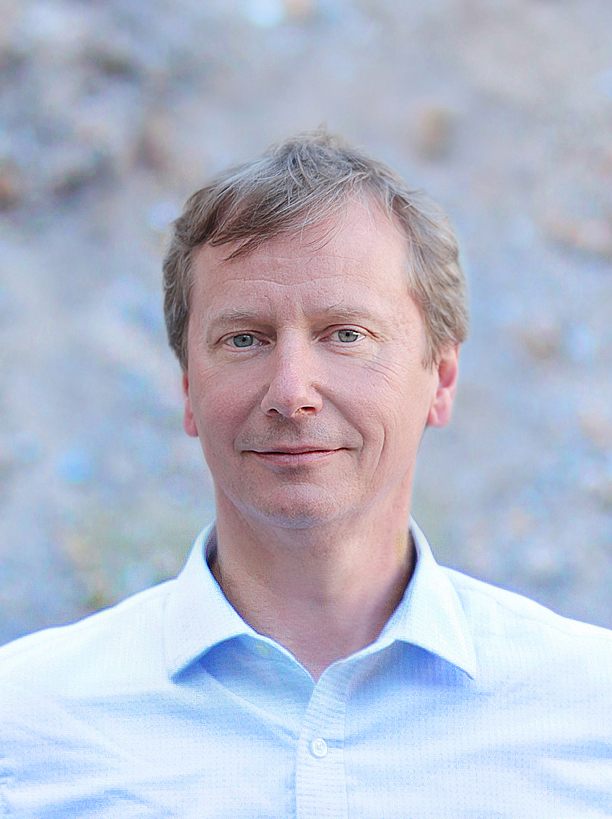}}]{Gerhard Fettweis}
	is Vodafone Chair Professor at TU Dresden since 1994, and heads the Barkhausen Institute since 2018, respectively. He earned his Ph.D. under H. Meyr's supervision from RWTH Aachen in 1990. After one year at IBM Research in San Jose, CA, he moved to TCSI Inc., Berkeley, CA. He coordinates the 5G Lab Germany, and 2 German Science Foundation (DFG) centers at TU Dresden, namely cfaed and HAEC. His research focusses on wireless transmission and chip design for wireless/IoT platforms, with 20 companies from Asia/Europe/US sponsoring his research.
	Gerhard is IEEE Fellow, member of the German Academy of Sciences (Leopoldina), the German Academy of Engineering (acatech), and received multiple IEEE recognitions as well has the VDE ring of honor. In Dresden his team has spun-out sixteen start-ups, and setup funded projects in volume of close to EUR 1/2 billion. He co-chairs the IEEE 5G Initiative, and has helped organizing IEEE conferences, most notably as TPC Chair of ICC 2009 and of TTM 2012, and as General Chair of VTC Spring 2013 and DATE 2014. 
\end{IEEEbiography}
\end{document}